\documentclass[twocolumn,trackchanges]{aastex62}

\newcommand{\msol}{M_\odot}
\newcommand{\rhoj}{\langle \rho_J \rangle}

\received{11 May, 2018}
\revised{12 July, 2018}
\accepted{12 July, 2018}

\submitjournal{ApJ}

\shorttitle{Cluster dissolution and centrally-peaked SFE profile}
\shortauthors{Shukirgaliyev et al.}

\begin{document}

\title{The long-term evolution of star clusters formed with a centrally-peaked star-formation-efficiency profile}

\correspondingauthor{Bekdaulet Shukirgaliyev}
\email{beks@ari.uni-heidelberg.de}

\author[0000-0002-4601-7065]{Bekdaulet Shukirgaliyev}
\altaffiliation{Fellow of the International Max Planck Research School for \\Astronomy and Cosmic Physics at the University of Heidelberg \\(IMPRS-HD)}
\affiliation{Zentrum f\"ur Astronomie der Universit\"at Heidelberg, Astronomisches Rechen-Institut, M\"onchhofstr. 12-14, 69120 Heidelberg, Germany}
\affiliation{Fesenkov Astrophysical Institute, Observatory str. 23, 050020 Almaty, Kazakhstan}

\author{Genevi\`{e}ve Parmentier}
\affiliation{Zentrum f\"ur Astronomie der Universit\"at Heidelberg, Astronomisches Rechen-Institut, M\"onchhofstr. 12-14, 69120 Heidelberg, Germany}

\author[0000-0002-5144-9233]{Andreas Just}
\affiliation{Zentrum f\"ur Astronomie der Universit\"at Heidelberg, Astronomisches Rechen-Institut, M\"onchhofstr. 12-14, 69120 Heidelberg, Germany}

\author{Peter Berczik}
\affiliation{National Astronomical Observatories of China and Key Laboratory for
Computational Astrophysics, Chinese Academy of Sciences, 20A Datun Rd,
Chaoyang District, 100012 Beijing, China}
\affiliation{Main Astronomical Observatory, National Academy of Sciences of Ukraine,
27 Akademika Zabolotnoho St, 03680 Kyiv, Ukraine}

\begin{abstract}

We have studied the long-term evolution of star clusters of the solar neighborhood, starting from their birth in gaseous clumps until their complete dissolution in the Galactic tidal field. 
We have combined the ``local-density-driven cluster formation model'' of \citet{PP13} with direct N-body simulations of clusters following instantaneous gas expulsion. 

We have studied the relation between cluster dissolution time, $t_{dis}$, and cluster ``initial'' mass, $M_{init}$, defined as the cluster  {mass at the end of the dynamical response to gas expulsion (i.e. violent relaxation), when the cluster age is 20-30 Myr}.
We consider the ``initial'' mass to be consistent with other works which neglect violent relaxation. 
The model clusters formed with a high star formation efficiency (SFE -- i.e. gas mass fraction converted into stars) follow a tight mass-dependent relation, in agreement with previous theoretical studies.
However, the low-SFE models present a large scatter in both the ``initial'' mass and the dissolution time, and a shallower mass-dependent relation than high-SFE clusters. 
 {Both groups differ in their structural properties on the average. }
Combining two populations of clusters, high- and low-SFE ones, with domination of the latter, yields a cluster dissolution time for the solar neighborhood in agreement with that inferred from observations, without any additional destructive processes such as giant molecular cloud encounters.

An apparent mass-independent relation may emerge for our low-SFE clusters when we neglect low-mass clusters (as expected for extra-galactic observations), although more simulations are needed to investigate this aspect.

\end{abstract}


\section{Introduction} \label{sec:intro}
The dissolution of star clusters has been the topic of many works. \citet{BL03} proposed an empirical law of cluster disruption, in which the cluster disruption time depends on the cluster initial mass (aka Mass Dependent Dissolution or MDD).
It was supported by both observational \citep[e.g.][]{Lamers+05a, Lamers+05b, Bastian+12, Silva-Villa+14} and theoretical works \citep[e.g.][]{Baumgardt+2002,BM03,GB08,Lamers+2010}.

However, \citet{Whitmore+07}, analyzing the cluster population of the Antennae galaxy-merger, proposed that during the first Gyr of evolution, the dissolution time of star clusters is not only independent of their mass but also of their environment.
This led them to propose the empirical ``universal law'' of cluster dissolution \citep[aka Mass Independent Dissolution or MID,][]{Whitmore2017}. Although their MID scenario was supported by several follow-up observational studies \citep[e.g. ][]{Fall+09,Chandar+10,FallChandar12,Chandar+14,Chandar+16,Linden+17} no theoretical work was able to support it, until \citet{Ernst+15} showed, by means of $N$-body simulations, that MID is a potential channel of cluster dissolution during the first Gyr of cluster evolution.

\citet{Ernst+15} performed direct $N$-body simulations of star clusters with different  {initial} Roche-volume filling factors, including clusters overfilling their Roche lobe (i.e. overfilling clusters).
They found that the  {dissolution time of the latter is independent} of their mass.
In their simulations, star clusters are gas-free and are initially in virial equilibrium.
\citet{Ernst+15} argued that clusters can overfill their Roche volume as a result of residual star-forming gas expulsion, i.e. when the gas is removed from the embedded cluster by stellar feedback, thereby weakening the gravitational potential.
Therefore, the Jacobi radius of gas-free clusters shrinks and the stars which inhabit the cluster outskirts can now be located beyond the new Roche volume.
 However, straight after gas expulsion star clusters are not in virial equilibrium as suggested in \citet{Ernst+15}.
  Instead, star clusters expand after gas expulsion if they were in equilibrium with the residual gas potential \citep{BK2007,Brinkmann+17,Bek+17}.
  The degree of spatial expansion is especially high for those with a low global star-formation efficiency (SFE, i.e. the star-forming gas mass fraction converted into stars).
 {
	\citet{Goodwin2009} argued that the critical factor for a cluster to survive gas expulsion is the dynamical state of stars (as measured by the virial ratio) immediately before gas expulsion, rather than SFE. 
	If $Q_0$ is the virial ratio of a cluster with its gas component, then the virial ratio of stars after instantaneous gas expulsion is $Q_\star = Q_0/\epsilon$ if the stars and gas follow the same shape for their respective density profiles (here $\epsilon$ is the SFE).
	\citet{Goodwin2009} reported that clusters with $Q_\star<1.5$ (i.e. $\epsilon>0.33$ if $Q_0=0.5$) are able to survive instantaneous gas expulsion. 
	However if the stars were \textit{not in virial equilibrium} with the gaseous content before gas expulsion, then the SFE needed to survive instantaneous gas expulsion will also be different. 
	Therefore, the effective SFE $\epsilon_{\mathrm{e}}$ (the SFE derived from the virial ratio of stars $\epsilon_{\mathrm{e}}=\displaystyle{\frac{1}{2Q_\star}}$) is the right parameter to measure survivability of the cluster after instantaneous gas expulsion. 
	Our  model clusters with a global SFE (i.e. `true' SFE) of $SFE_{gl}=0.15$ define the limit of post-gas-expulsion survivability and actually have a virial ratio of about $Q_\star=1.55$ \citep[see Table 1 of][]{Bek+17}, which is only slightly higher than that needed to survive according to \citet{Goodwin2009}. 
	However, the non-equality of the `true' SFE ($SFE_{gl}$) and the effective SFE ($\epsilon_\mathrm{e}$) in our models is not caused by the non-virial equilibrium state of the embedded cluster before gas expulsion as discussed by \citet{Goodwin2009}.  
	Rather, it is the consequence of stars having a density profile steeper than the gas immediately before gas expulsion (i.e the stars are more concentrated towards the center than the gas) while being in virial equilibrium with the gas potential. More details about our initial conditions can be found in \citet{Bek+17}. 
    Our model clusters with $SFE_{gl}=0.15$ can expand so much after gas expulsion, that they become overfilling-cluster candidates, due to the fact, that they survive instantaneous gas expulsion with only slightly higher virial ratio than that required to survive instantaneous gas expulsion.}

In our previous work \citep{Bek+17} we have studied how star clusters respond to instantaneous gas expulsion when they form according to the local-density-driven cluster formation model of \citet{PP13}.
 That is, model clusters form in centrally-concentrated spherically-symmetric molecular clumps with a constant star-formation efficiency per free-fall time.
  As a consequence, their \textit{stellar volume density profile is steeper than that of the initial and residual star-forming gas}.

In this work, we expand the results obtained in  our previous paper \citep{Bek+17} and we look at the problem of cluster dissolution during the first Gyr of evolution anew.
 In \citet{Bek+17} we have studied the violent relaxation  of star clusters for different stellar masses and global SFEs.
 Now we focus on their long-term evolution. We also focus on our model clusters formed with a low global SFE, i.e. $SFE_{gl}=0.15$, and investigate if they behave in a similar way to the overfilling cluster models of \citet{Ernst+15}, and if they show evidence of MID.
 In contrast to their study, our model clusters bear the information about their formation conditions and the violent relaxation which follows gas expulsion.

In section \ref{sec:models} we describe our cluster models, simulations and  cluster mass estimators we have used in this study.
In section \ref{sec:stoch} we discuss the stochasticity effects in low-SFE model clusters.
We investigate how the evolution of our model clusters depend on their initial mass and central density in section \ref{sec:mddvsmid}.
Finally, in section \ref{sec:conc} we provide our conclusions.
\section{Methods} \label{sec:models}
\subsection{Cluster models}
 In \citet{Bek+17} we have performed a set of direct $N$-body simulations to  study the response of star clusters to instantaneous gas expulsion.
 We have considered clusters with different birth stellar masses (i.e. the stellar mass of a cluster at the time of instantaneous gas expulsion, $M_\star =$ 3k, 6k, 10k, 15k, 30k $\msol$), different global SFEs (i.e the star-forming gas mass fraction converted into stars before gas expulsion, $SFE_{gl} = 0.10, 0.13, 0.15, 0.20, 0.25$), and mean volume densities (quantified by the half-mass radius to Jacobi radius ratio $r_h/R_J=0.025, 0.05, 0.07, 0.1$ of the gas-free cluster).
 We have assumed that star clusters form according to the local-density-driven cluster formation model of \citet{PP13} (\textit{i.e. stars have a density profile steeper than that of the residual gas at the time of gas expulsion}). 
  {However, while \citet{PP13} infer the density profile of the residual gas and of the embedded cluster starting from that of the initial gas, we infer the density profile of the initial and residual gas from the assumed Plummer density profile of the embedded cluster.
 Therefore, in all model embedded clusters, stars follow a Plummer profile at the time of gas expulsion.
 Initial and residual gas, do not, however, and are different for different SFEs \citep[see Fig. 2 of][]{Bek+17}.
 We recover the total and residual gas density profiles using the cluster formation model of \citet[][see their Eqs. 19-20]{PP13} \citep[for more details see Sec. 2 of][]{Bek+17}}. 
 We have used the program \textsc{mkhalo} from the \textsf{falcON} package \citep{McMillan2007} to generate the initial conditions of our $N$-body models assuming that embedded clusters are in virial equilibrium with the total gravitational potential (i.e stars + gas) at the time of gas expulsion.
 The considered model clusters are on a circular orbit in the Galactic disk plane at a Galactocentric distance of $R_G=8$ kpc.
 The Galactic tidal field is given analytically by an axisymmetric three-component Plummer-Kuzmin model \citep{MiyamotoNagai75} with the parameters as given in \citet{Just+09} (their Eq.~(32) and their Table 1).
 We assume that all stars reach the zero age main sequence at the time of instantaneous gas expulsion for simplicity, therefore the time-span since gas expulsion equals the cluster age in our simulations.
 The initial mass function (IMF) of \citet{K2001} with initial star mass limits of $m_\mathrm{low} = 0.08\ M_\odot$ and $m_\mathrm{up} = 100\ M_\odot$ has been applied. 
 For more information about the initial conditions and the violent relaxation phase of cluster evolution, we refer our reader to \citet{Bek+17}.
  {By ``violent relaxation'' we mean the evolution of gas-free clusters from a super-virial state into a new state of quasi-equilibrium following instantaneous gas expulsion. 
 We assume that the violent relaxation ends when model clusters stop (violently) losing their mass in response to gas expulsion as a result of an initial state of non-equilibrium. 
 We note, however, that our model clusters may not have fully re-virialized at that time (i.e., the end of violent relaxation, as we define it, does not coincide exactly with the time of cluster re-virialization). 
 In this paper we refer to the end of violent relaxation as $t=30\text{ Myr}$ after instantaneous gas expulsion.}

 Here, in this study, we continue our existing set of direct $N$-body simulations until the full dissolution of the model star clusters in the tidal field of the Galaxy.
 We use only the \citet{Bek+17}' models with $r_h/R_J=0.05$ and which survive as bound clusters after violent relaxation (i.e. $SFE_{gl}\geq 0.15$). We completed this initial model set with newly-run $SFE_{gl}=0.17$ models for some birth masses and $SFE_{gl}=0.15$ models for birth masses higher than what was considered in \citet{Bek+17} (i.e $M_\star=60\text{k }\msol$~and~$100\text{k }\msol$, equivalent to $N_\star = 105 554$~and~$174 257$ stars, respectively, for a \citet{K2001} IMF with $m_\mathrm{low} = 0.08\ M_\odot$ and $m_\mathrm{up} = 100\ M_\odot$).
 All model clusters considered in this study have evolved from the time of instantaneous gas expulsion until full dissolution.
 Therefore all of them bear the information about their formation conditions and violent relaxation.
 The full parameter space covered by our $N$-body simulations is provided in Table \ref{tab:models}. 
 All our high-resolution direct $N$-body simulations were performed using the \textsc{$\phi$grape-gpu} code developed by \citet{Berc2011} with single-stellar evolution recipes of \textsc{sse} by \citet{Hurley+00}. 
 We have run our simulations on high-performance computing clusters: \textsc{jureca} \citep[][ J\"ulich, Germany]{jureca}, \textsc{laohu} (NAOC/CAS, Beijing, China), \textsc{kepler} (ARI/ZAH, Heidelberg, Germany) and GPU part of \textsc{bwForCluster: mls\&wiso} (Heidelberg Unversity, Heidelberg, Germany).

\begin{deluxetable}{r|r|r|r|l|l}
\tablewidth{0pt}
\tablecolumns{6}

\tablecaption{Model cluster parameters. The columns are as follow: (1) total number of stars, (2) birth mass, (3) global SFE, (4) number of random realizations per model (number of those calculated till cluster dissolution), (5) mean bound mass fraction at the end of violent relaxation and its standard deviation when more than one random seed, (6) mean dissolution time. \label{tab:models}}
\tablehead{ 
\colhead{$N_\star$} & \colhead{$\displaystyle\frac{M_\star}{\msol}$} & \colhead{$SFE_{gl}$} & \colhead{$n_{rnd}$} & \colhead{$\langle F_{bound}\rangle$} & \colhead{$\displaystyle\left\langle\frac{t_{dis}}{10^9\text{ yr}} \right\rangle$} 
}
\colnumbers
\startdata
  5225 &   3000 & 0.15 & 21 (21) &0.07$\pm$0.05 &	$0.19\pm0.11 $ \\ 
  5225 &   3000 & 0.17 &  1 ( 1) &0.21 &		$0.52$ \\ 
  5225 &   3000 & 0.20 &  1 ( 1) &0.29 &		$0.71$ \\ 
  5225 &   3000 & 0.25 &  1 ( 1) &0.53 &		$1.07$ \\ 
 10455 &   6000 & 0.15 & 26 (26) &0.06$\pm$0.04 &	$0.23\pm0.16 $ \\ 
 10455 &   6000 & 0.17 &  1 ( 1) &0.21 &		$0.66$ \\ 
 10455 &   6000 & 0.20 &  3 ( 1) &0.31$\pm$0.02 &	$1.29 $ \\ 
 10455 &   6000 & 0.25 &  3 ( 1) &0.50$\pm$0.03 &	$1.56 $ \\ 
 17425 &  10000 & 0.15 & 26 (26) &0.06$\pm$0.03 &	$0.33\pm0.15 $ \\ 
 17425 &  10000 & 0.20 &  1 ( 1) &0.38 &		$2.11$ \\ 
 17425 &  10000 & 0.25 &  1 ( 1) &0.51 &		$2.55$ \\ 
 26138 &  15000 & 0.15 & 22 (22) &0.08$\pm$0.04 &	$0.53\pm0.24 $ \\ 
 26138 &  15000 & 0.17 &  1 ( 1) &0.18 &		$1.02$ \\ 
 26138 &  15000 & 0.20 &  3 ( 1) &0.33$\pm$0.02 &	$1.84 $ \\ 
 26138 &  15000 & 0.25 &  3 ( 1) &0.51$\pm$0.01 &	$2.86 $ \\ 
 52277 &  30000 & 0.15 & 16 (16) &0.08$\pm$0.03 &	$0.64\pm0.32 $ \\ 
 52277 &  30000 & 0.17 &  1 ( 1) &0.16 &		$1.37$ \\ 
 52277 &  30000 & 0.20 &  1 ( 1) &0.34 &		$3.58$ \\ 
 52277 &  30000 & 0.25 &  1 ( 1) &0.56 &		$5.49$ \\ 
104554 &  60000 & 0.15 & 15 ( 9) &0.08$\pm$0.02 &	$0.75\pm0.35 $ \\ 
174257 & 100000 & 0.15 &  3 ( 3) &0.08$\pm$0.01 &	$0.96\pm0.34 $ \\ 
\enddata
\end{deluxetable}
\subsection{Random realizations}
 In \citet{Bek+17} we mentioned that the bound mass fraction at the end of violent relaxation, $F_{bound}$ (i.e. the stellar mass fraction remaining bound to the cluster at an age of 20-30 Myr), can vary by about 6-10 percent of the birth mass for different random realizations of a given model.
 Such variations are almost negligible for model clusters with a high $SFE_{gl}>0.15$, as they survive with more than 20 percent of the birth mass.
 However, the situation changes for our $SFE_{gl}=0.15$ models, as they survive with the lowest bound mass fraction (about or even below 0.10), being the transition models from full destruction following gas expulsion to survival in the parameter space of $SFE_{gl}$.
 For such a low bound mass fraction the variations mentioned above are significant.
 Therefore, we performed additional random realizations of the $SFE_{gl}=0.15$ models to have better statistics for our study. 
 In forth column of Table \ref{tab:models} we give the numbers of random realizations per model and in parentheses the number of runs completed till cluster dissolution.
 The column 5 shows the mean bound mass fractions and their standard deviations measured at $t=30\text{ Myr}$.
 The discussions and results on random realizations are presented further in section \ref{sec:stoch}.

\subsection{Cluster mass estimates}

 To estimate cluster masses is a complex issue, both from an observational and theoretical point-of-view.
 Even the very definition of a star cluster varies through the literature \citep[e.g. see review in ][]{Renaud18}.
 To estimate the luminous mass of an observed star cluster is not straightforward due to incompleteness issues and field star contamination.
 To estimate the dynamical mass of an observed cluster is hindered by the contribution of binaries to the cluster overall velocity dispersion.
 From a theoretical point of view, the mass determination is not straightforward due to the unknown second integral of motion.

 In \citet{Bek+17}, we refer to the cluster mass as the stellar mass within one Jacobi radius, $R_J$, which is also supposed to be the bound mass.
 To calculate the Jacobi radius of a cluster at a given age, we start with its value at the time of gas expulsion and with the stellar mass it contains using Eq (13) from \citet{Just+09}.
 Then we re-calculate the Jacobi radius using the mass within the previously defined Jacobi radius.
 We iterate until the Jacobi radius converges.
 Our Jacobi radius calculation method works only if we define the cluster density center correctly, which can prove an issue for the following reasons.
 In our $N$-body simulations we keep track of all stars, even those which have definitively escaped the cluster in which they initially formed.
 Therefore, escaped stars live in the tidal tails of our model clusters, which can extend as far as to wrap around the Galaxy making it difficult to define the cluster center as a center of mass of all stars.
 That the tidal tails contain epicyclic over-densities yields difficulties to find the exact cluster center too.
 That is, we have struggled to find the correct cluster center in the late stages of cluster evolution, as they are becoming low-mass and diffuse objects.
 Their low densities can then be comparable to that of the surrounding field, or to the epicyclic over-densities of the tidal tails.
 Our algorithm can thus incorrectly identify the center of a tidal tail over-density as the cluster center.
 In order to prevent this, we use to calculate the cluster center only those stars which were within $2 R_J$ of it in the previous $N$-body simulation snapshot.
 This allowed us to steer clear of the epicyclic over-densities of the tidal tails.
 Yet, it remains difficult to define correctly the cluster center, when it consists of few stars, has an extended core, a low volume density, or presents sub-structures.
 Usually, this happens in the last stages of evolution of star clusters, before they get fully destroyed by the Galactic tidal field.
 In such clusters, relatively small shifts of the assumed cluster center can lead to significantly wrong mass estimates because of a wrong Jacobi radius, $R_J$.

In order to avoid this, we assume that clusters are dissolved if they have less than 100~$\msol$ left within 2 Jacobi radii:
\begin{equation}
\label{eq:tdis}
t_{dis}=t(M_{2J}<100 M_\odot),
\end{equation}
where $M_{2J}$ is the stellar mass enclosed within 2 Jacobi radii.

 We have checked how long some of our model clusters can live beyond the dissolution time, $t_{dis}$, defined by Eq. (\ref{eq:tdis}).
 Typically the difference is about a few tens of Myr, which is not significant, especially for clusters whose lifetime scales up to a Gyr.

\begin{figure*}[ht!]
\plotone{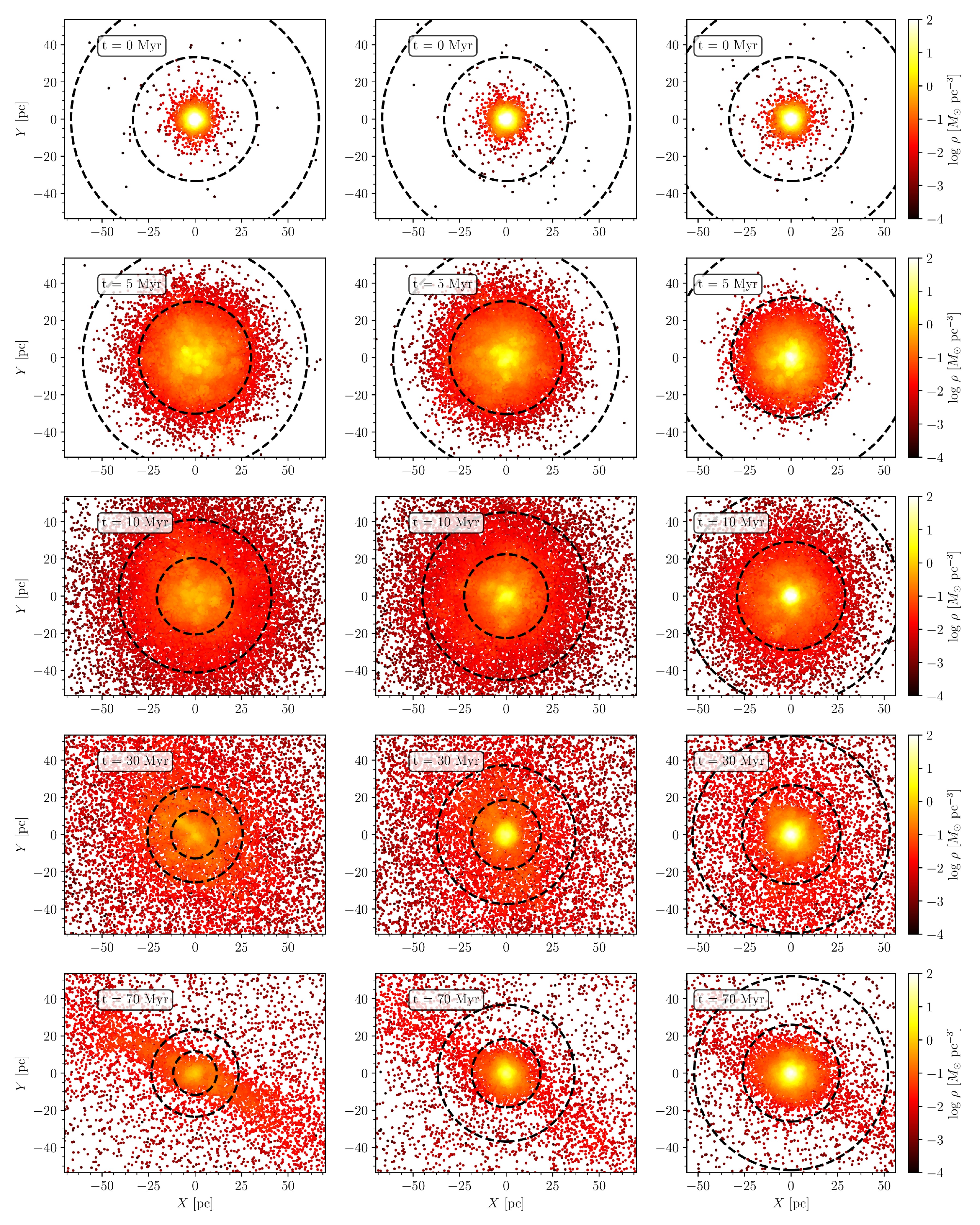}
\caption{Volume density maps of clusters with a birth mass of $M_\star=15000\msol$ projected onto the Galactic plane. 
The left and middle columns correspond to two random realizations of a model cluster with $SFE_{gl}=0.15$, and the right column corresponds to a $SFE_{gl}=0.25$ model. 
From top to bottom, we provide 5 different snapshots of each model cluster  at times $t=0,5,10,30,70$~Myr. 
Each point corresponds to one star whose color-coding depicts the local volume density calculated by means of a 50-nearest-neighbor scheme. {\footnotesize\textit{ {Note that the color-scale does not show densities higher than 100 $\msol \mathrm{pc}^{-3}$ in order to highlight the color contrast in low density regions at a later time of cluster evolution. The central densities at the time of gas expulsion are in fact as high as $1.6\cdot10^3\msol \mathrm{pc}^{-3}$.}}} 
The dashed circles correspond to $R_J$ and $2R_J$. 
The bound fractions at $t=30\text{ Myr}$ are, from left to right: $F_{bound}=0.06, 0.18 \text{ and } 0.5$. The corresponding dissolution times are $t_{dis}=0.3\text{ Gyr, } 1.2 \text{ Gyr and } 2.9 \text{ Gyr}$, respectively.\label{fig:densevol}}
\end{figure*}

In this contribution we consider two types of cluster-mass estimates.
 One is the Jacobi mass (or ``bound mass''), $M_J$, which is the stellar mass within one Jacobi radius, $R_J$.
 The other one, which we refer to as the ``extended mass'' $M_{2J}$, is the stellar mass within $2R_J$.
 The second mass estimate is important for young clusters as they are surrounded by an envelope of unbound stars \citep{EFF87}, most of them located beyond one Jacobi radius but still within $2R_J$.
 Such envelopes persist for many Myrs, as Fig. \ref{fig:densevol} will show.
 They can count towards the mass of clusters in extra-galactic studies, where a membership analysis is impossible, and the cluster mass is estimated by fitting the cluster surface brightness profile.

 Figure \ref{fig:densevol} visualizes the evolution of star clusters with a birth mass of $M_\star=15\text{k }\msol$ in the form of volume density maps.
 Each point in these plots correspond to the position of one star projected onto the Galactic disk plane.
 The colors correspond to the local volume density obtained by a nearest-neighbor scheme with $N_{nb}=50$ neighbor stars.
 The coordinate system is centered on the density center of  the clusters and the two circles correspond to $1R_J$ and $2R_J$.
 The left and middle panels depict two random realizations of the model with $SFE_{gl}=0.15$ and the right column corresponds to $SFE_{gl}=0.25$.
 In each column, different snapshots are presented ($t=0,5,10,30,70$~Myr). 
 
\begin{figure*}[ht!]
	\plotone{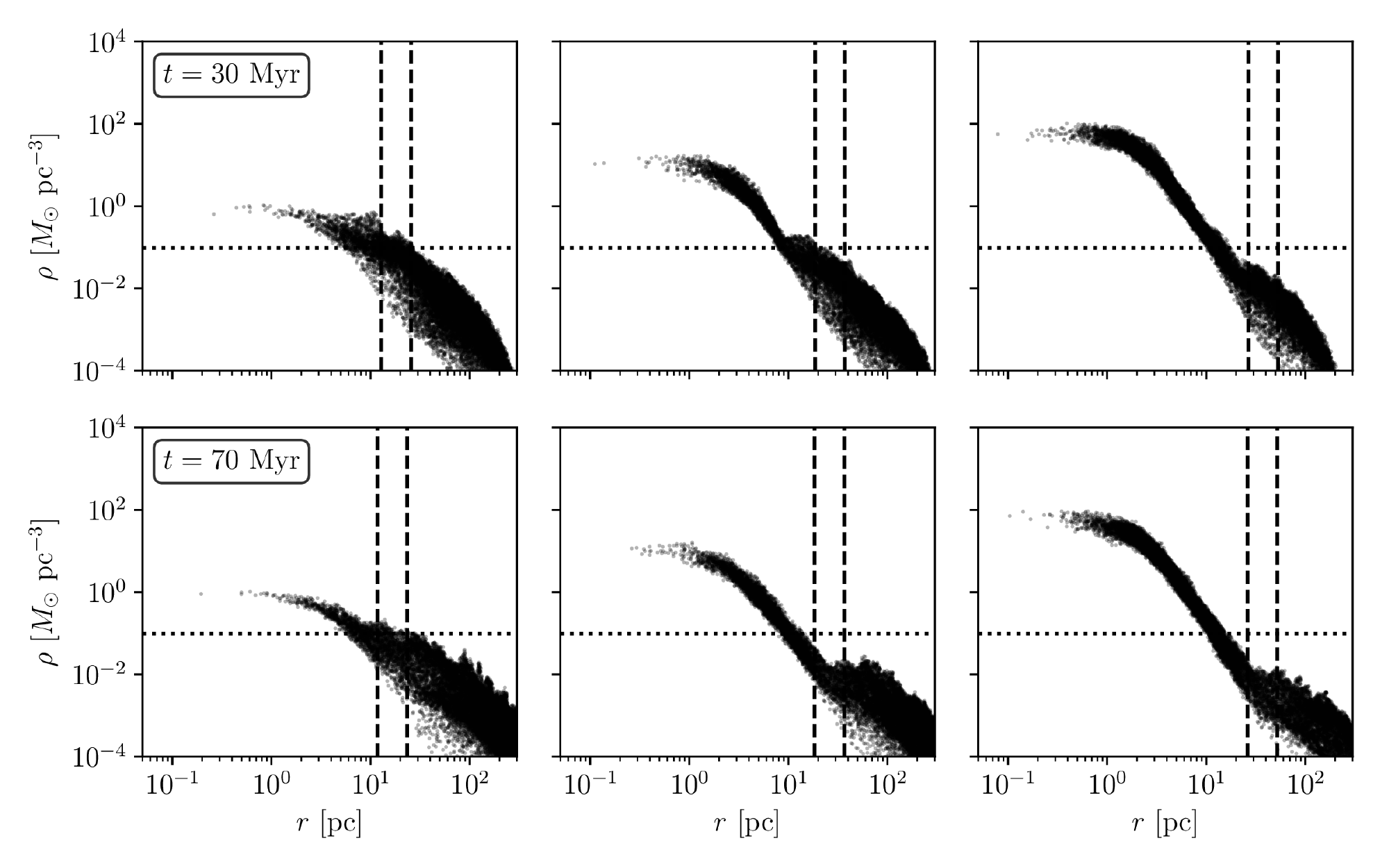}
	\caption{Volume density profiles of model clusters whose birth mass is $M_\star=15000\msol$ (same models as in Fig. \ref{fig:densevol}). Each point represents the density at the location of one star. Left and middle panels correspond to two random realizations of the $SFE_{gl}=0.15$ model, while the right panels correspond to the $SFE_{gl}=0.25$ model cluster. The top and bottom panels correspond to the density profiles calculated at $t=30$~Myr and  $t=70$~Myr, respectively. In each panel, the vertical lines show the location of $1R_J$ and $2R_J$. The horizontal lines correspond to the mean density within one Jacobi radius, $\rhoj$, which is constant for the considered Galactic orbit of star clusters.
	\label{fig:dens}}
\end{figure*}

 In Fig. \ref{fig:dens} we show the radial volume density profiles of the same 3 model clusters as in Fig. \ref{fig:densevol} 
calculated at the end of violent relaxation ($t=30$~Myr, top panels), and at a later time when clusters are almost cleared of their envelope stars ($t=70$~Myr, bottom panels).
 Each point represents the density at the location of one star, calculated using a 50-nearest-neighbor scheme.
 In each panel, the vertical dashed lines correspond to 1 and 2 Jacobi radii, and the horizontal dotted line corresponds to the mean stellar density within one Jacobi radius, i.e.
$ \rhoj=\displaystyle{M_J}/\left({\displaystyle4/3\pi R_J^3}\right)$.
According to Eq. (13) of \citet{Just+09}, for a given environment, i.e. for a fixed circular orbit in the Galactic disk plane, the mean density within one Jacobi radius, $\rhoj$, is constant and independent of cluster parameters, since $R_J\propto M_J^{1/3}$. In our case, for a circular orbit with $R_G=8$~kpc, the mean density within one Jacobi radius is $\rhoj\approx 0.1 \msol \text{pc}^{-3}$.

 In the left panels of Fig. \ref{fig:dens}, the region between $1R_J$ and $2R_J$ is well populated by stars such that the distant observer, when calculating the cluster mass by fitting its projected density profile, will use the envelope stars too.
 For the middle and right columns, however, the envelope stars do not contribute much to the cluster mass, as evidenced by the density contrast between the central part and the outskirts of the clusters (see Fig. \ref{fig:densevol} and Fig. \ref{fig:dens}).
 These density contrasts are about 1-2 orders of magnitude in the left panels, about 3 orders of magnitude in the middle panels and about 4-5 orders of magnitude in the right panels.

\section{Stochasticity during cluster expansion}\label{sec:stoch}
 
\begin{figure}[!h]
\plotone{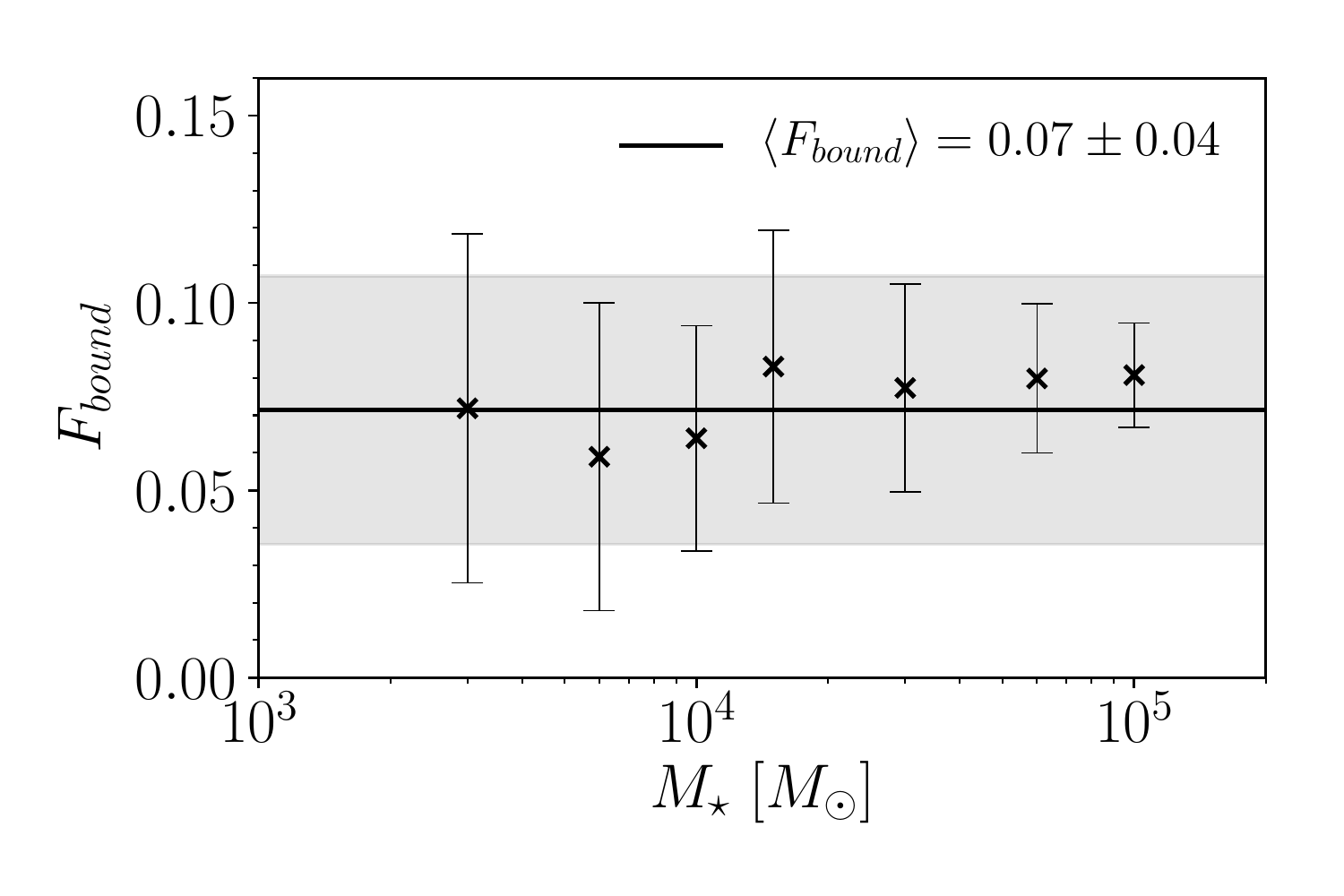}
\caption{The mean bound mass fractions at the end of violent relaxation obtained from random realizations of cluster models with $SFE_{gl} = 0.15$ as a function of the birth mass of star clusters. The error-bars correspond to the standard deviations. The solid line shows mean bound mass fraction of all model clusters with $SFE_{gl}=0.15$ and shaded area corresponds to the standard deviation.\label{fig:meanbf}}
\end{figure}

 Based on our random realizations of $SFE_{gl}=0.15$ model clusters, we have studied how distributed the bound mass fractions $F_{bound}$ at the end of violent relaxation ($t=30\text{ Myr}$) are. $F_{bound}$ is here defined as the ratio between the Jacobi mass and the birth mass of a cluster: 
 $$ F_{bound} = M_J(t=30\text{Myr})/M_\star. $$
 Figure~\ref{fig:meanbf} shows the mean bound mass fractions 
 of our $SFE_{gl}=0.15$ cluster models as a function of the birth masses, where the error-bars depict the standard deviation. 
 The solid line and the shaded area depict the total mean bound mass fraction, $\langle F_{bound} \rangle$, and total standard deviation, obtained for all $SFE_{gl}=0.15$ model clusters. 
 We find $\langle F_{bound} \rangle = 0.07\pm0.04$. 
 As we see from Fig. \ref{fig:meanbf} the mean bound mass fraction does not significantly depend on the birth mass, being equal to the total mean bound mass fraction within the error bars.
 That the standard deviations are decreasing with increasing birth mass can result from the smaller numbers of random realizations implemented for models with higher birth mass (see the column 4 in Table \ref{tab:models} for $SFE_{gl}=0.15$ models).
 We also find that the bound mass fractions do not change significantly from $t=30 \text{ Myr} $ to $t=70\text{ Myr}$.
 
  Our simulations show that a number of stars as high as $N_\star=100\text{k}$ ($M_\star=60\text{k}\msol$) does not remove the relatively large scatter characterizing the bound fraction of the $SFE_{gl}=0.15$ models.
   We think that this might be a consequence of the Poisson noise in the initial phase-space distribution of stars, and of some stochasticity taking place during the cluster expansion.
  {Our model clusters with a global SFE as low as ${SFE}_{gl} = 0.15$ can expand so much after gas expulsion that their central density drops down significantly. 
 Density sub-structures can form inside the Jacobi radius -- where the density profile also becomes shallower -- as a result of the local gravitational potential wells of stars more massive than $8 M_\odot$.}
 Even during the early expansion phase, individual high-mass stars can attract and retain many co-moving stars in their vicinity. 
 That is, stars co-moving with a nearby high-mass star can get trapped by its gravitational potential.
 As a result, these surrounding stars start to move collectively towards their neighboring high-mass star and deepen the local potential well. This collective motion continues even after the high-mass star goes supernova. 
  The more high-mass stars involved in this sub-cluster formation process, the more massive the sub-cluster formed.
 Eventually, the sub-structures formed during expansion can merge into one bigger cluster, or expand further and dissolve depending on their bound mass and kinetic energy.
 This can be seen by comparing the left and middle panels of Fig. \ref{fig:densevol}, as we discuss later in this section.

 Stochastic effects could be significant at this stage due to the relatively small number of massive stars which do not escape the cluster.
 Since we have applied the \citet{K2001} IMF, the number fraction of massive stars is about 0.6 percent.
 That means we have about 600 massive stars at the time of gas expulsion for a cluster with $N_\star=100$k stars and about 60 high-mass stars for a $N_\star=10$k cluster.
 If about 93 percent of these massive stars escape to the field, as expected for a $SFE_{gl}=0.15$ model whose mean bound fraction is $F_{bound}=0.07\pm 0.04$, at the end of violent relaxation we are left with about 42 massive stars within one Jacobi radius $R_J$ for the former, and about only 4-5 high-mass stars for the latter.

 If we look at the left and middle columns of Fig. \ref{fig:densevol} we can see the evolution of two initially identical cluster models with $M_\star=15\text{k } \msol$ and $SFE_{gl}= 0.15$.  
 Although model clusters in the left and middle panels are almost identical at the time of gas expulsion ($t=0\text{ Myr}$), they slightly differ from each other already at $t=5$~Myr, the one in the middle panel being slightly more centrally-concentrated than the other.
 Both clusters have almost the same Jacobi radii $R_J$, and therefore, the same bound mass (i.e. the stellar mass within one Jacobi radius, $R_J$).
 However, this is still the expansion phase.
 The middle-panel cluster has a slightly higher number of massive stars close to its center, while the massive stars of the left-panel cluster are distributed broadly within $R_J$. 
 This is seen by the distribution of local over-densities at the centers of which high-mass stars are usually located. 

 Later, at $t=10$~Myr the difference becomes even clearer.
  {In the left-panel, the region occupied by the local potential wells driven by the high-mass stars stretch almost to the Jacobi radius.
 In the middle-panel, this region is more centrally located.}
 Again, at this time, both clusters have comparable bound masses, but markedly different structures already.

 At the end of violent relaxation at $t=30$~Myr (when clusters stop losing mass in response to gas expulsion, although not the re-virialization time yet) and later on at $t=70\text{ Myr}$, these two $SFE_{gl}=0.15$ clusters present markedly different bound masses, even though they started with the same birth mass and the same global SFE. 
 This is, the consequence of the highly-stochastic spatial distribution of high-mass stars.
 
 By $t=70$~Myr model clusters have re-virialized and regain a more spherically-symmetric shape within the Jacobi radius, which becomes a good estimator of the cluster radius onward. 
  {That is the clusters are now cleaned from the most of envelope stars and we can see the tidal tails as streaky features at $t=70\text{ Myr}$ (see the lowest panels of Fig. \ref{fig:densevol}). 
 We remind the reader that we use an axisymmetric Galactic potential (bulge, halo, disk), which does not include any features such as spiral arms, bar or disk wrap.}

\section{What does cluster dissolution depend on?}\label{sec:mddvsmid}
\subsection{Cluster life expectancy and initial cluster mass} \label{sec:sc_life}
\begin{figure}[!ht]
\plotone{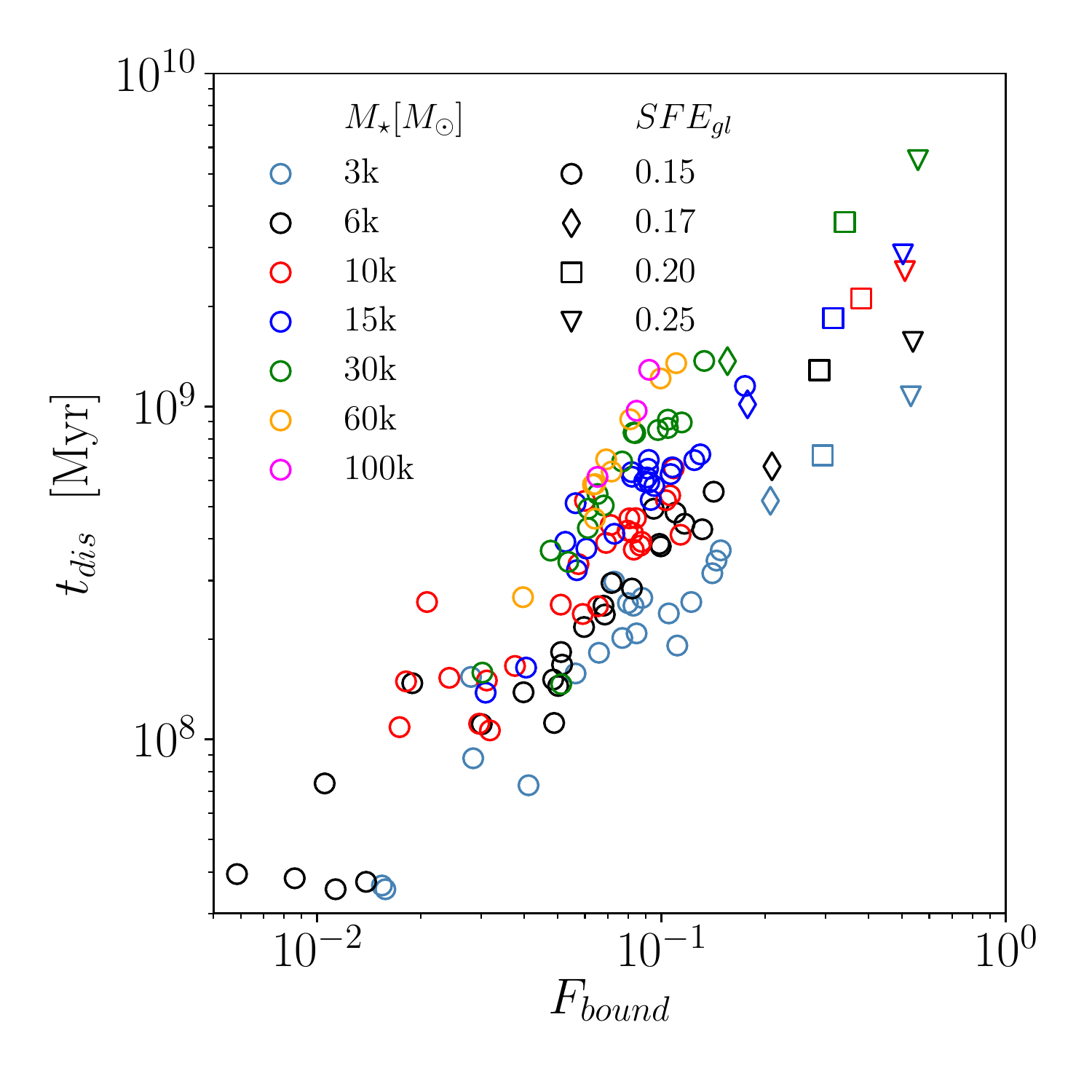}
\caption{Cluster dissolution time as a function of bound mass fraction. Cluster birth mass and global SFE are indicated by colors and symbols, respectively, according to the key.\label{fig:tdisFb}}
\end{figure}

 In this section we study the relation between cluster dissolution time and mass. 
 From the random realizations of model clusters with $SFE_{gl}=0.15$ we find that the differences in bound mass fraction at the end of violent relaxation also results in different star cluster lifetimes.
 Figure \ref{fig:tdisFb} presents the dissolution time $t_{dis}$ of star clusters as a function of the bound mass fraction $F_{bound}$ at the end of violent relaxation.
 The cluster dissolution time is defined according to Eq. (\ref{eq:tdis}).
 The color-coding and symbol-coding correspond to cluster birth mass and global SFE, respectively (see the key).
 The general trend is that the higher the bound mass fraction, the longer lives a cluster. 
However, the large scatter and the fact that Fig.~\ref{fig:tdisFb} considers various birth masses do not give us much more information about the relation between cluster dissolution time and its mass.

\begin{figure*}[ht!]
\plotone{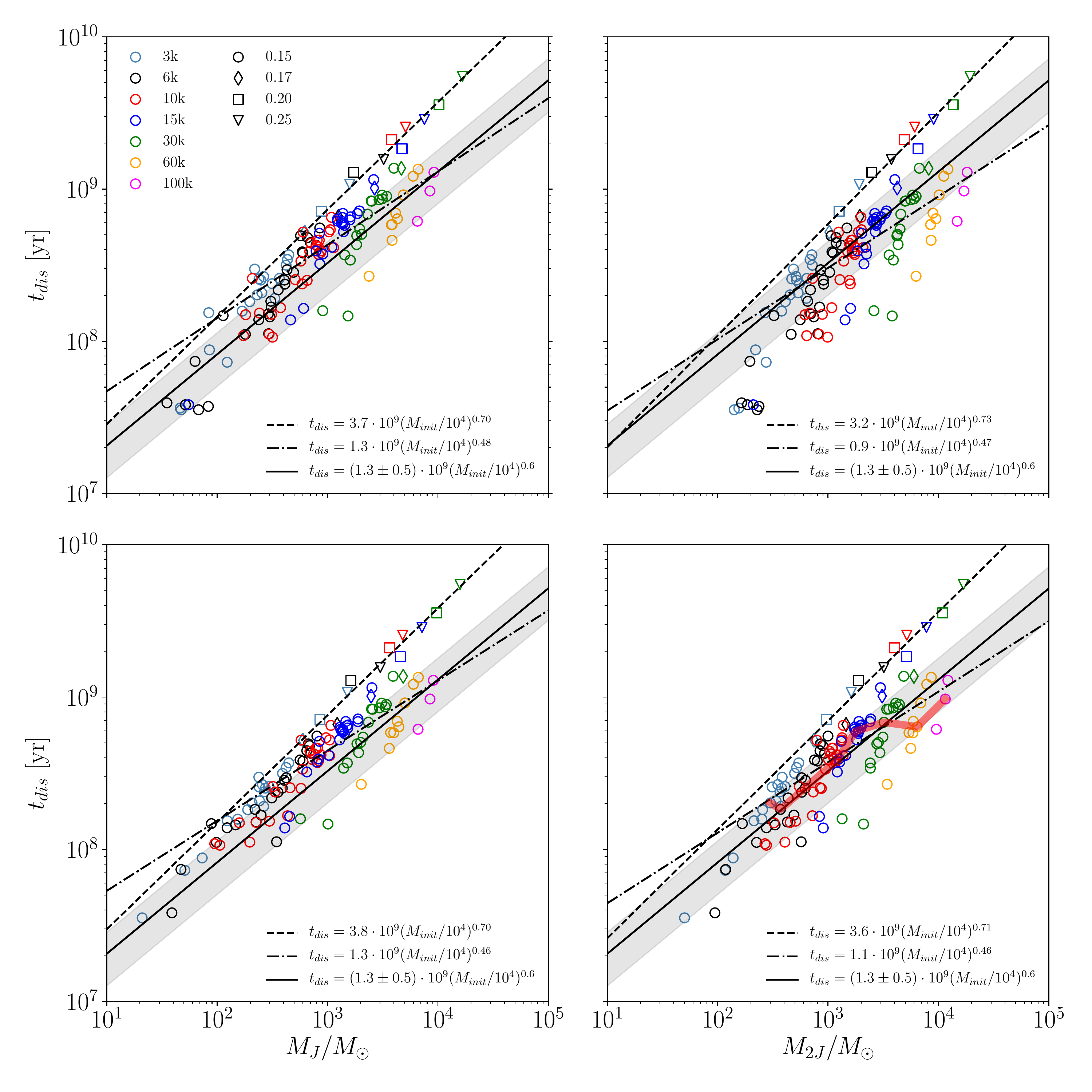}
\caption{Star cluster dissolution time versus cluster mass at the end of violent relaxation ($t=30$~Myr in top panels and $t=70$~Myr in bottom panels).
We refer to this as the cluster ``initial'' mass. 
 The left panels show the cluster ``initial'' mass defined as the Jacobi mass, $M_J$, while the right panels present it as the ``extended mass'', $M_{2J}$.
Each point represents one cluster model, with the color-coding defining the cluster birth mass, and marker shapes coding the global SFE.
The solid line with shaded area corresponds to the cluster disruption model for the solar neighborhood of \citet{Lamers+05a}.
The dashed  and dash-dotted lines depict the best fits to high-SFE ($SFE_{gl}\geq0.20$) and low-SFE ($SFE_{gl}=0.15$) model clusters. 
 The red curve in the lower-right panel connects the median random realizations of the models with $SFE_{gl}=0.15$.\label{fig:lifemass}}
\end{figure*}

The cluster dissolution time, $t_{dis}$, as a function of cluster ``initial'' mass is presented in Fig. \ref{fig:lifemass}.
 By ``initial'' mass, $M_{init}$, we mean here the cluster mass once violent relaxation is over, that is, when the long-term secular evolution starts.
 Our cluster ``initial'' masses are therefore measured at $t=30$~Myr (top panels) and $t=70$~Myr (bottom panels), and are lower than the birth masses given in Table \ref{tab:models} (the ratio between Jacobi and birth masses is the bound fraction $F_{bound}$).
  We have done so to be consistent with studies which ignore the violent relaxation phase of cluster evolution when inferring the cluster dissolution time as a function of cluster mass  {(for instance by considering only clusters older than 10-30 Myr)}.

 When plotting the cluster dissolution time as a function of cluster initial mass, not only do we 
 measure the mass at two different cluster ages,
 we also consider two definitions of cluster masses in terms of cluster spatial coverage.
 Specifically, the cluster initial mass is defined either as the Jacobi mass $M_J$ (left panels), or as  the ``extended mass'' $M_{2J}$ (right panels).
 We therefore investigate 4 different cases in total.
 The color-coding shows the birth mass of the model clusters, while the different symbols correspond to different $SFE_{gl}$ (see the key).
 The scatter arising from the random realizations of a given model (given birth mass and given $SFE_{gl}$) is therefore illustrated by symbols of a given color and of a given type.
 
The solid line, with the shaded area accounting for the error-bars, corresponds to the MDD relation of \citet{BL03} 
 \begin{equation}
 t_{dis} = t_4^{dis} \left(\displaystyle\frac{M_{init}}{10^4\msol} \right)^{\gamma},
 \end{equation}
 where $M_{init}$ is cluster ``initial'' mass and $t_4^{dis}$ is the dissolution time for a cluster with ``initial'' mass of $M_{init}=10^4\msol$.
 Here we re-call that $M_{init}$ is the equivalent of the Jacobi mass, $M_J$, or the extended mass, $M_{2J}$, at $t=30\text{ or } 70\text{ Myr}$, but not of the birth mass, $M_\star$.
 The values of $t_4^{dis}=(1.3\pm0.5)\cdot10^9\text{ Gyr}$ and $\gamma=0.6$ are taken from \citet{Lamers+05a} for the solar neighborhood.
 The dashed and dash-dotted lines show the best fits to our high-SFE ($SFE_{gl}\geq0.20$) and low-SFE ($SFE_{gl}=0.15$) model clusters, respectively. 
 The bold red curve in the lower-right panel connects, for each cluster birth mass (i.e. for each symbol color), the medians of $SFE_{gl}=0.15$ model random realizations. 

 The overall impression from Fig.~\ref{fig:lifemass} is that star clusters dissolve in agreement with MDD, although with some significant scatter.
 Especially model clusters formed with a relatively high global SFE ($SFE_{gl}\geq0.20$, open squares and triangles in Fig.~\ref{fig:lifemass}) nicely follow an MDD relation (dashed line), with a slope a bit steeper ($\gamma\sim0.7$) than that given by \citet{BL03}, and a dissolution time longer  {($t_4^{dis}\sim3.8\text{ Gyr}$, lower left panel of Fig. \ref{fig:lifemass})} than the  {observational} estimate of \citet{Lamers+05a} for the Solar Neighbourhood (solid line).
  {Our result} is consistent with other theoretical works, that Roche-volume-filling or under-filling clusters in virial equilibrium dissolve in a mass-dependent way.
 However, the cluster dissolution time obtained from our high-SFE models ($t^{dis}_4 \sim 3.8 \text{ Gyr}$) is almost a factor of two shorter than that predicted by \citet{BM03} {, who found} $t^{dis}_4 = 6.9 \text{ Gyr}$.
 This can result from different models for the Galactic gravitational potential: while we consider an axisymmetric three-component model \citep{Just+09}, \citet{BM03} consider a spherically-symmetric logarithmic potential.
 Another probably more crucial reason is the ``initial'' density profile of the star clusters: our density profiles at $t=30$~Myr differ from the virialized King models (with $W_0=5\text{ and }W_0=7$) used by \citet{BM03}.

 We do not notice any significant difference for the high-SFE models between the 4 panels. Therefore, for the high-SFE models, how the cluster initial mass is defined (at $t=30\text{ or } t=70 \text{ Myr;}$ bound mass, $M_J$ or extended mass, $M_{2J}$) hardly influences the corresponding predicted MDD relation.

 In contrast to high-SFE ones, low-SFE models show broad scatter and differ from panel to panel.
 The scatter is maximum for extended mass at $t=30\text{ Myr}$ (upper right panel) and minimum for Jacobi mass at $t=70\text{ Myr}$ (lower left panel).
 Since the difference between the Jacobi mass and the extended mass is larger for low-SFE clusters and negligible for high-SFE clusters the initial masses are characterized by a broader scatter in right panels than in left panels of Fig.~\ref{fig:lifemass}.
 When the clusters have evolved a bit more in time, the envelope gets cleaned up by an age of $t=70$~Myr and the extended mass of our model clusters then becomes comparable to their Jacobi mass (bottom panels of Fig.~\ref{fig:lifemass}){, although the scatter persists. }
 Our best fits to low-SFE models provide a dissolution time in agreement with \citet{Lamers+05a}, $t_4^{dis} \sim 1.3 \text{ Gyr}$, but a slightly shallower slope $\gamma\sim0.5$  {(lower left panel of Fig. \ref{fig:lifemass}).}
 For the $SFE_{gl}=0.15$ models, we can see that the scatter in bound mass fractions can yield significant differences in the cluster lifetime (open circles of a given color). 
As shown in Fig~\ref{fig:densevol} and \ref{fig:dens}, (left and middle panels) this is the result of the development with time of markedly different density profiles. 
 Additionally, if we consider a vertical bin embracing cluster masses from $2\times 10^3\text{ to }4\times 10^3\msol$ in the right panels of Fig.~\ref{fig:lifemass},
 we can see that such clusters can dissolve as fast as within 100--200 Myr or can live as long as about a Gyr.

\begin{figure}[!h]
\plotone{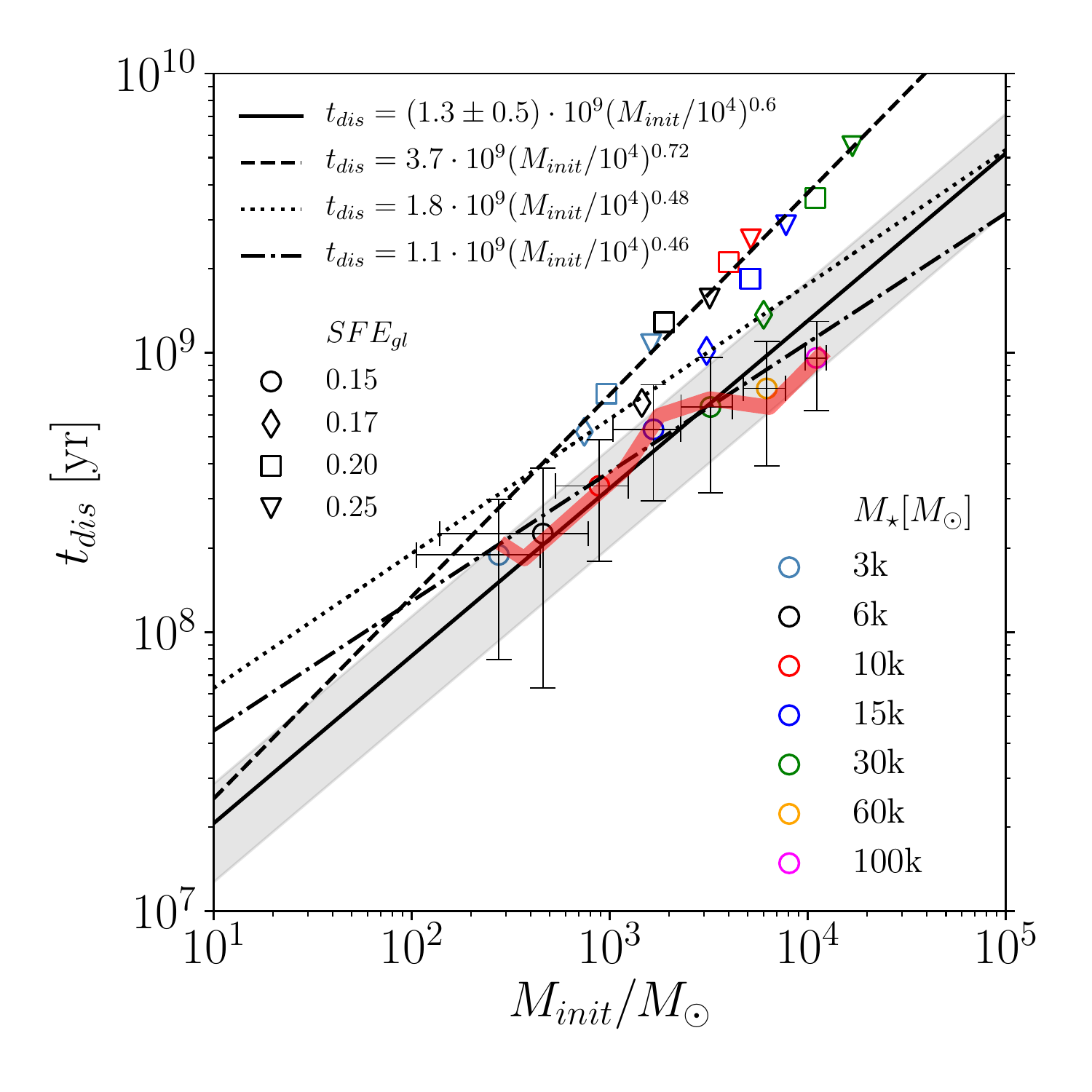}
\caption{ Cluster dissolution time as a function of extended mass at $t=70\text{ Myr}$. The $SFE_{gl}=0.15$ models are represented by the mean extended mass and mean dissolution time per model with error-bars representing the standard deviations. The red curve connects median random realizations of each model as in the lower right panel of Fig. \ref{fig:lifemass}. The dashed, dotted and dash-dotted lines are best fits to the $SFE_{gl}\geq0.20$, $SFE_{gl}=0.17$ and $SFE_{gl}=0.15$ models. The solid line with shaded area corresponds to the MDD relation of \citet{BL03} for the solar neighborhood \citep{Lamers+05a}. \label{fig:meantdis}}
\end{figure}

  For the sake of clarity Fig.~\ref{fig:meantdis} shows each low-SFE model (i.e. given birth mass and global SFE) represented by its mean and standard deviation.
The initial mass is defined as the extended mass, $M_{2J}$, and the age is $t=70\text{ Myr}$.
Figure~\ref{fig:meantdis} is thus equivalent to the lower right panel of Fig.~\ref{fig:lifemass}, apart from the low-SFE models being represented by mean values.
 Now the results for the $SFE_{gl}=0.17$ models can be seen clearly as open diamonds with the corresponding fit having $t_4^{dis}=1.8\text{ Gyr}$ and $\gamma\sim0.5$ (dotted line). 
The other lines are as in the lower right panel of Fig.~\ref{fig:lifemass}, namely, the MDD relation of \citet{BL03} and the best fits to the $SFE_{gl}\geq 0.20$ and $SFE_{gl}=0.15$ models.
The combination of these 3 groups of star clusters can yield a relation close to that of \citet{BL03}, especially if low-SFE clusters dominate the cluster population.  
 This is a real possibility for the solar neighborhood, since observations of nearby gas-embedded clusters tell us that low-SFE embedded clusters are more common than high-SFE ones \citep{LL03,Evans+09,Peterson+11,KE12,Kainulainen+14}. 
 
 \citet{Gieles+06} noted that the dissolution time of a $10^4\msol$ cluster in the solar neighborhood as inferred from observations by \citet{Lamers+05a} ($t_4^{dis}=1.3\pm0.5\text{ Gyr}$) differs by about a factor of 5 from what was predicted by \citet{BM03} ($t_4^{dis}=6.9\text{ Gyr}$). They suggested that the discrepancy can be {eliminated} by accounting for the influence on star clusters of Giant Molecular Cloud encounters.
 Here, we propose yet another channel to explain the shorter dissolution time of \citet{Lamers+05a}, namely, that the star clusters of the solar neighborhood are predominantly the survivors of embedded clusters formed with a global SFE of $SFE_{gl}\approx0.15$. 
 We expect that most of these clusters will dissolve by the time they reach an age of 1~Gyr.
 In contrast to other theoretical studies who consider compact clusters in virial equilibrium as initial conditions, our model clusters have experienced violent relaxation, which is a natural process affecting the evolution of young clusters. 
 With our approach we are thus able to simulate the evolution of clusters which have survived violent relaxation as bound, but diffuse, objects. They dissolve faster than their compact counterparts, even for otherwise equal ``initial'' masses, and we probably observe them as open clusters in the solar neighborhood.   
 
\begin{figure}[!h]
\plotone{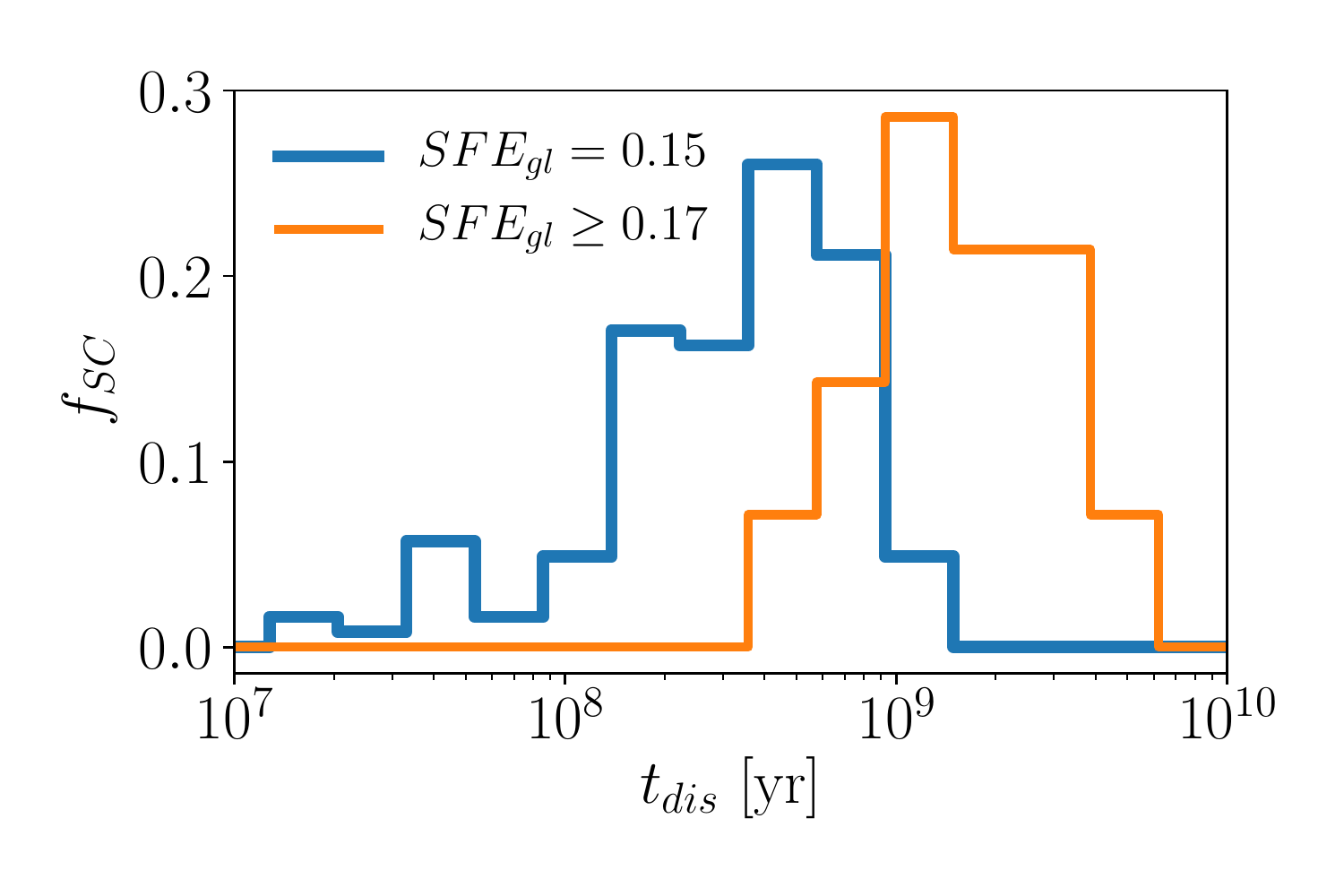}
\caption{Histogram of model clusters by dissolution times for low-SFE and high-SFE clusters (blue and orange lines, respectively). The area subtended by each histogram is unity.
 \label{fig:tdishist}}
\end{figure}

  If the low-SFE clusters dominate the cluster census of Galactic disk, then not many clusters are able to live beyond 1~Gyr as we can see from Fig. \ref{fig:tdishist}.  
 It shows the histogram of cluster dissolution times for low-SFE ($SFE_{gl}=0.15$, blue) and higher-SFE ($SFE_{gl}\geq0.17$, orange) clusters. 
 Each histogram is normalized such that the sum of all bins equals to unity.
 The combination of both distributions, each with each own contribution, will provide the distribution of dissolution times of a cluster population. 
 The latter can give us a hint about the shape of the corresponding cluster age distribution.
 A cluster population dominated by low-SFE clusters should feature a peak in cluster  logarithmic age distribution ($dN/d\log t$) earlier than 1 Gyr (if the cluster formation rate is constant), since most low-SFE clusters die before 1 Gyr. 
  The peak at about a few hundred Myr in the age distribution of solar neighborhood star clusters has been discussed in \citet{Lamers+05a} and \citet{Piskunov+18}. 

 {An interesting question is whether the dissolution time distribution of low-SFE clusters extends significantly beyond 1 Gyr, if our simulations were to include clusters more massive than those modeled so far.
Here we remind the reader that the initial cluster mass function is a power-law with a slope of $-2$, i.e. $\displaystyle\frac{dN}{dm}\propto m^{-2}$, equivalent to $\displaystyle\frac{dN}{d\log m} \propto m^{-1}$.
This implies that for a mass spectrum with a lower mass limit of $100\msol$, star clusters more massive than $10^4\msol$ ($10^5\msol$)
represent 1\% (0.1\%) of the cluster census.}

 {In addition, to produce a low-SFE cluster with $M_{init}=10^5\msol\ (10^6\msol)$ we need a molecular clump with a gas mass of $10^7\msol\ (10^8\msol)$. 
  This is because the ``initial'' mass of such clusters is only about 1\% ($F_{b}\times SFE_{gl} = 0.07\times0.15$) of the total mass of the star-forming clump.
  In our Galaxy very massive low-SFE clusters should thus be very rare, if they exist at all, because the mass function of molecular clumps is also a decreasing power-law with an index of $-1.7$ to $-2$.}

 {Finally, according to \citet{Rahner+17} the higher the mass of a star-forming clump the higher the SFE it requires to be destroyed by its newly-formed star cluster.
 Should it not be the case, the re-collapse of the stellar-feedback driven gaseous shell may happen and produce more stars, enhancing the SFE \citep{Rahner+18}.
  Therefore the distribution presented in Fig. \ref{fig:tdishist} is robust and will not change significantly if more massive low-SFE clusters were included. 
  }

An interesting trend emerges when we consider $SFE_{gl}=0.15$ models in Fig.~\ref{fig:meantdis}.
 If we ignore the low-mass end ($\leq10^3\msol$, i.e. the mass range for which the cluster sample is incomplete in most extra-galactic studies) and if we take into account that, in extra-galactic observations, the masses of young clusters can be over-estimated due to the contribution of an envelope of unbound stars (i.e. the cluster ``initial'' mass has to be defined as our extended mass, $M_{2J}$), an apparent MID mode may be emerging. 
 For initial mass $M_{2J}$ between $10^3$ and $10^4\msol$ low-SFE clusters actually show similar mean (and median) dissolution times, with even increasing standard deviations.
 On top of that, if we consider model clusters formed with very low $SFE_{gl}<0.15$, which do not survive instantaneous gas expulsion and the resulting violent relaxation ($t_{dis}<30\text{ Myr}$), but  {which are} still observable as young clusters, then the effect of an apparent MID can be further strengthened.

However, since we do not cover that large a range of cluster masses and do not have that high a number of random realizations for good statistics (especially at high-mass), we cannot argue firmly in favor of a mass-independent dissolution relation for low-SFE star clusters.
 Here, we stress that with a mean bound fraction at the end of violent relaxation of $F_{bound}= 0.07$, to model a star cluster with an ``initial'' mass $>10^4\msol$ and which formed with $SFE_{gl}=0.15$, requires a birth mass of at least $10^5\msol$.
 We are currently expanding our sample for $M_\star=10^5\msol$.
  {The modeling of even more massive clusters is highly-desirable for comparison with the results of observational extragalactic studies.}
 Nevertheless, we can firmly say that we have found a large scatter in the relation between the cluster ``initial'' masses and the cluster dissolution times for low-SFE clusters and for a given birth mass.
 This scatter results from the massive-star driven stochastic effects taking place during violent relaxation.
 Such effects yield, for a given birth mass and given $SFE_{gl}=0.15$, different density profiles (hence different degrees of cluster compactness) and different bound fractions at the end of violent relaxation (hence different cluster ``initial'' masses.

In summary so far, we have found that clusters formed with a high $SFE_{gl}$ ($>0.15$) dissolve in an MDD regime.
 Clusters formed with a low $SFE_{gl}$ ($=0.15$) also dissolve in an MDD regime, albeit with a significant scatter. Their dissolution time is comparable to that observationally inferred by \citet{Lamers+05a}.
There is a strong mass-dependent upper limit to the cluster dissolution time, which means that, for a given environment, low-mass clusters cannot live as long as their high-mass \textit{compact} counterparts.
 However, some high-mass clusters can dissolve as quickly as low-mass ones in the same environment.

\subsection{Cluster life expectancy and cluster central density}
\begin{figure*}
\plotone{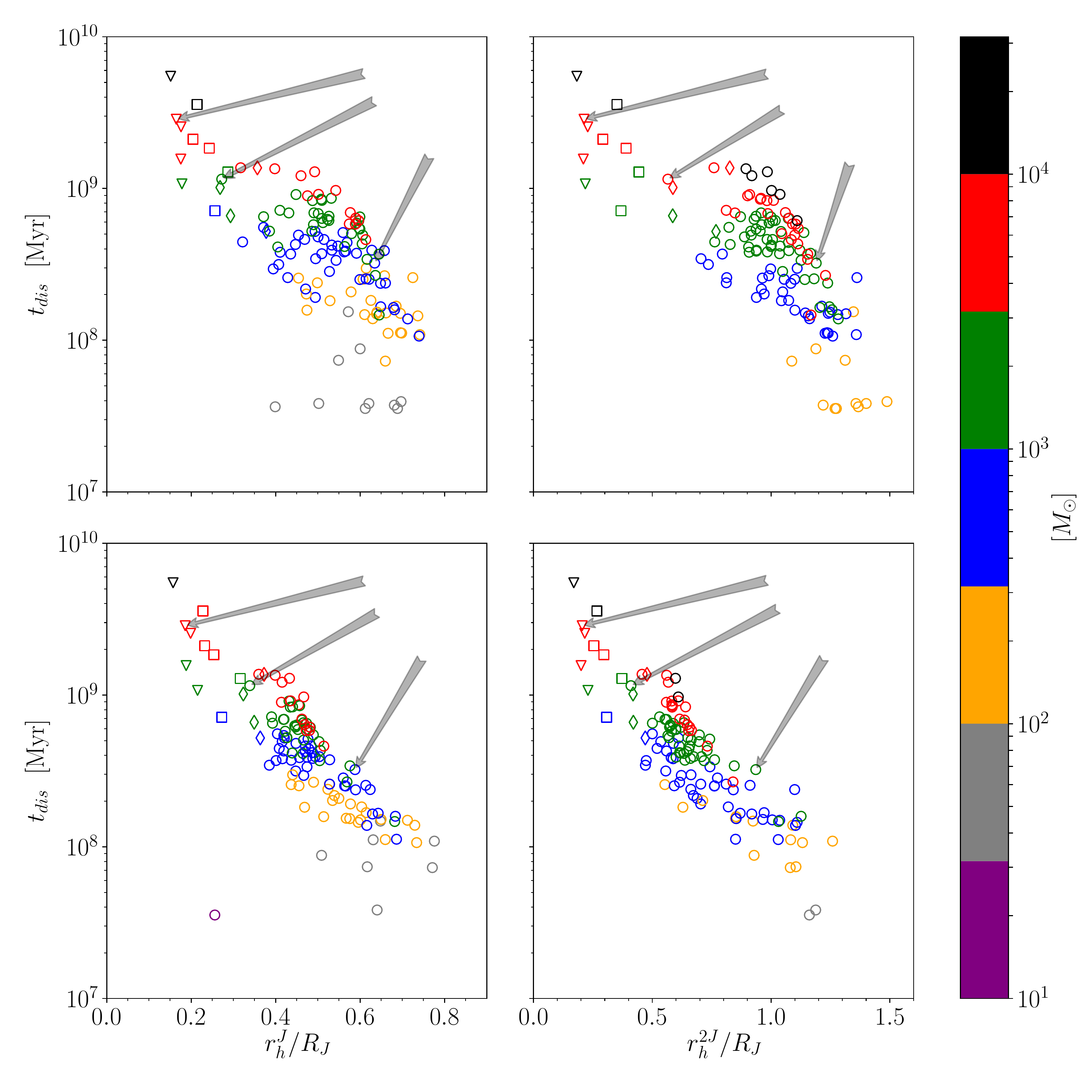}
\caption{Cluster dissolution time as a function of Roche volume filling factor at $t=30$ and $t=70\text{ Myr}$ (top and bottom panels, respectively).
In the left panels, the cluster initial mass is estimated as the Jacobi mass, $M_J$, while in the right panels it is estimated as the extended mass, $M_{2J}$.
Cluster initial masses are shown by the color-coding presented on the right-hand-side color-bar.
The half-mass radius, $r_h$, is measured as the radius containing half of the cluster initial mass and marked as $r_h^J$ when $M_{init}=M_J$ and $r_h^{2J}$ when $M_{init}=M_{2J}$. 
The gray arrows indicate those $M_\star=15\text{k }\msol$ model clusters presented in Figs.~\ref{fig:densevol} and \ref{fig:dens}.
 \label{fig:rhrt}}
\end{figure*}

Investigating further the parameters of our model clusters we have found a correlation between the cluster dissolution time, $t_{dis}$, and the Roche volume filling factor measured at the end of violent relaxation, although with significant scatter (Fig.~\ref{fig:rhrt}).
The Roche volume filling factor is defined as the half-mass to Jacobi radius ratio, and the half-mass radius refers to the Jacobi mass or the extended mass (left and right panels of Fig. \ref{fig:rhrt}, respectively). 
 The filling factors are calculated at $t=30\text{ and }70\text{ Myr}$ in top and bottom panels as in Fig. \ref{fig:lifemass}. 
 The color-coding  depicts the Jacobi mass, $M_J$ (left panels) and the extended mass, $M_{2J}$ (right panels) indicated by the common color-bar at the right-hand-side.
 Note that the color coding is different from Fig.~\ref{fig:lifemass} where it refers to the birth mass of clusters.
 The shapes of markers still show the global SFEs.
{ With gray arrows we indicate those $M_\star=15000\msol$ clusters presented in Figs.~\ref{fig:densevol} and \ref{fig:dens}, namely from right-to-left, two random realizations of the $SFE_{gl}=0.15$ model (open circles) and one random realization of $SFE_{gl}=0.25$ model (open triangle).
 The indicated model clusters here are in the reverse order to that of the order of panels in Fig.~\ref{fig:densevol}.}

 The correlation is such that for a given ``initial'' mass, the higher the filling factor, the shorter the dissolution time, as found in \citet{Ernst+15}.
 However, a comparison between our results and those of \citet{Ernst+15} is not fully self-consistent.
 Firstly, our model clusters are still expanding and have not returned to virial equilibrium yet when they over-fill their Roche volume. 
 In contrast, model clusters of \citet{Ernst+15} are initially in virial equilibrium while overfilling their Roche volume.
 Secondly, our star clusters can present different density profiles (shallow or steep, with extended or compact core) after violent relaxation, therefore the Roche volume filling factor, as defined here, cannot characterize them universally.

\begin{figure*}[ht]
\plotone{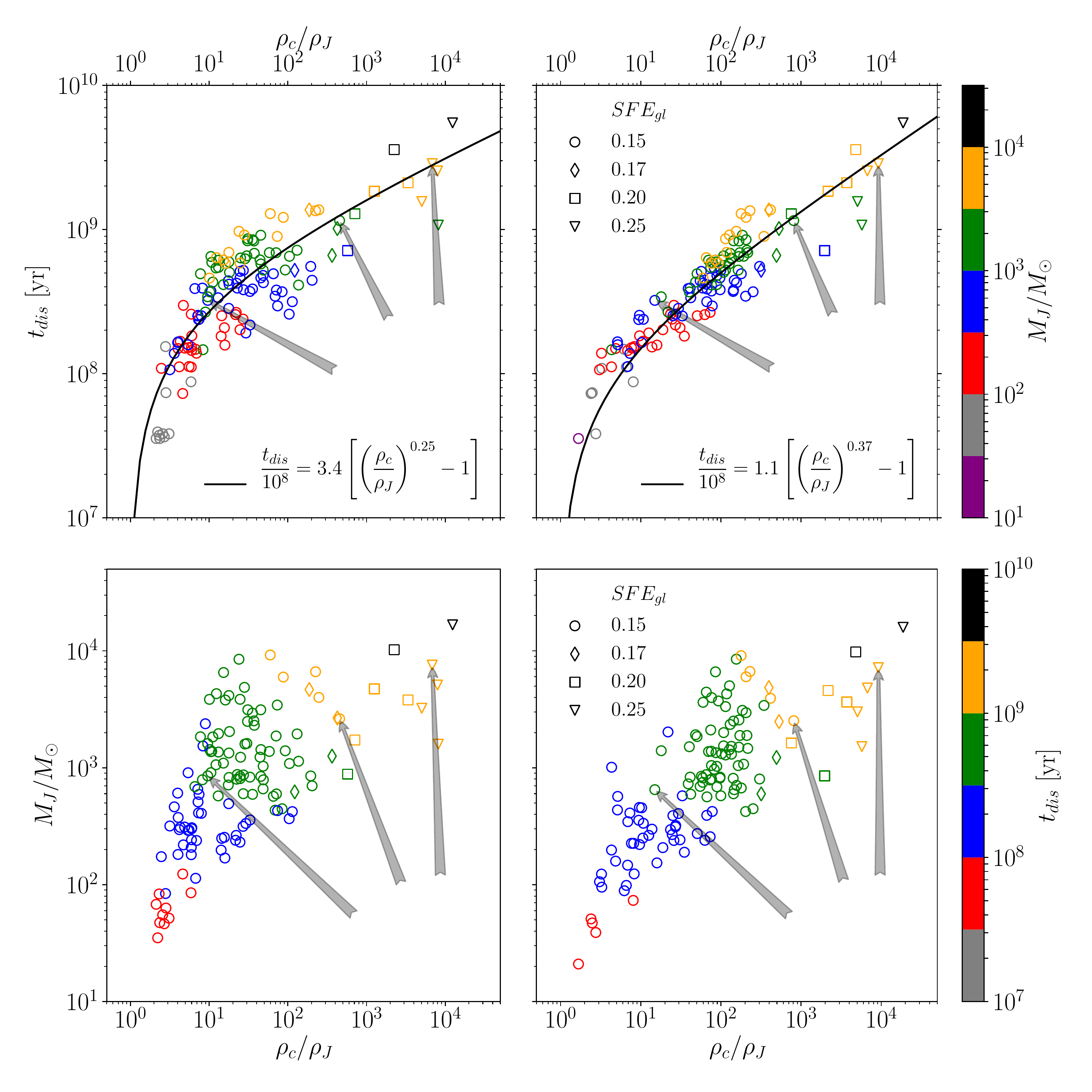}
\caption{Cluster dissolution time, $t_{dis}$ (top panels) and cluster Jacobi mass, $M_J$ (bottom panels), as functions of volume density contrast between cluster center and outskirts, $\rho_c/\rho_J$ (Eq.~\ref{eq:rhocrhoj}) measured at $t=30$~Myr and $t=70$~Myr (left and right panels, respectively).
Unlike in Fig.~\ref{fig:lifemass} the color-coding refers to the cluster Jacobi mass at quoted ages in top panels, and to the cluster dissolution time in bottom panels. 
 {In top panels, the best fits for the cluster dissolution time }(Eq. \ref{eq:rhofit1}-\ref{eq:rhofit2}) are shown with black curves. 
The three $M_\star=15000\msol$ model clusters presented in Fig.~\ref{fig:densevol} and Fig.~\ref{fig:dens} are indicated with the arrows in each panel in the same order, i.e. from left to right, as the corresponding panels of Fig.~\ref{fig:densevol} and Fig.~\ref{fig:dens}.
\label{fig:rho}}
\end{figure*}

 We also find a correlation between the cluster life expectancy, $t_{dis}$, and the volume density contrast between cluster center and outskirts, defined as the central volume density, $\rho_c$, to the volume density at Jacobi radius, $\rho_J$, ratio: 
 \begin{equation}
 \rho_c/\rho_J=\displaystyle\frac{\rho(r=0)}{\rho(r=R_J)}.
 \label{eq:rhocrhoj}
 \end{equation}
 This is shown in upper left panel of Fig. \ref{fig:rho} as measured at the end of violent relaxation ($t=30$~Myr).
 The smaller the density contrast, the faster the cluster dissolves and vice versa.
 The correlation is the tightest if we consider the density contrast of slightly more evolved clusters ($t=70$~Myr, top right panel of Fig. \ref{fig:rho}), when the envelope at $R_J<r<2R_J$ has been cleaned up of most of its stars.
The color-coding corresponds to the Jacobi mass at the corresponding ages.
 The best fits, in the form of
 \begin{equation}
  t_{dis} = \displaystyle3.4\left[\left(\frac{\rho_c}{\rho_J}\right)^{0.25}-1\right]\times 10^8\text{ yr} \label{eq:rhofit1} 
 \end{equation}
 and
 \begin{equation}
 t_{dis} = \displaystyle1.1\left[\left(\frac{\rho_c}{\rho_J}\right)^{0.37}-1\right]\times 10^8\text{ yr}, \label{eq:rhofit2}
\end{equation}
are shown with black curves in the top left and top right panels respectively.
For high density contrast ($\rho_c/\rho_J>10^3$) clusters, the top panels of Fig.~\ref{fig:rho} show that their dissolution times are  mass-dependent, which is again consistent with earlier studies of dissolution of Roche Volume filling or under-filling clusters. 

 When a cluster has a shallow density profile, the shrinking of its Jacobi radius due to cluster mass-loss (e.g. stellar evolution, tidal stripping, etc.) will leave outside the new Jacobi radius a number of stars higher than in case of a steeper density profile.
 Such clusters therefore dissolve more quickly than those with a steep density profile.
 Alternatively, if a cluster has developed a large core, comparable to the Roche volume, this also leads to the total destruction of the cluster.
 Such clusters are located at the very left of the top panels in Fig.~\ref{fig:rho}, with low density contrasts and therefore short lifetimes.

In order to see better the correlation between these three parameters, $\rho_c/\rho_J,t_{dis}\text{ and }M_J$, we swapped the dissolution time for the Jacobi mass in the bottom panels. 
The bottom panels of Fig.~\ref{fig:rho} show the Jacobi mass, $M_J$, as a function of the density contrast, $\rho_c/\rho_J$, color-coded by the dissolution time, $t_{dis}$.
{As it is seen now clearly here with more or less nice pattern the dissolution time depends on both ``initial'' mass and density contrast, wherein short dissolution time of massive clusters are explained by their low density contrast (low concentration) and vice versa.
This is also consistent with other theoretical works, where the evolution of globular clusters with initially low concentration has been discussed (see e.g. \citet{FH95}, \citet{TnPZ00} and \citet{VZ03})   }

As we see from the foregoing results, when studying the long-term secular evolution of star clusters it is worth estimating their ``initial'' mass at $t=70$~Myr after gas expulsion  in order not to be biased by the processes taking place during violent relaxation. 
 The latter is very different form the subsequent long-term evolution of star clusters. 
 As we showed in \citet{Bek+17}, clusters formed with very low global SFE ($<0.15$) dissolve during violent relaxation, and their dissolution time is independent of their birth mass.
 
 We remind the reader that we have modeled the clusters in a given tidal-field environment (i.e. star clusters have the same mean density at the time of instantaneous gas expulsion), and that they have the same stellar density profile at the time of instantaneous gas expulsion. 
 All relations we have found could thus be affected if birth conditions are different.
 Star clusters which at the time of gas expulsion are more compact or more diffuse  than our models may have evolutionary tracks different from those analyzed here, forming denser or more diffuse bound clusters after violent relaxation.

\section{Discussions and Conclusions}\label{sec:conc}

 We have performed a large set of direct $N$-body simulations of the evolution of star clusters in the solar neighborhood starting from their birth in molecular clumps until complete dissolution in the Galactic tidal field.
 We have not considered any hydro-dynamical simulations to account for the formation of our model embedded clusters.
 Instead we have used a semi-analytical approach -- the local-density-driven cluster formation model of \citet{PP13}, and assumed instantaneous gas expulsion.
 The modeling of the evolution of our clusters therefore covers the time-span from their formation to complete dissolution.
As a result, our model clusters bear the information about their formation conditions and their violent relaxation all through their evolution, even for dissolution times longer than 1 Gyr.

 We have found that model clusters with $SFE_{gl}=0.15$ present a significant stochasticity during the expansion phase.
 That is, initially almost identical clusters can follow very different tracks in terms of bound mass fraction, structure and dissolution time.
 At the end of violent relaxation these clusters retain quite a small bound mass fraction, with relatively large scatter, $F_{bound}=0.07\pm0.04$. 
 The bound mass fraction $F_{bound}$ does not depend on the cluster birth mass (see Fig. \ref{fig:meanbf} for $SFE_{gl}=0.15$ models or column (5) in Table \ref{tab:models}) and the scatter persist even for a number of stars as high as $N_\star\approx174\text{k}$.
 The reason could be in the relatively high virial ratio of low-SFE clusters ($Q\sim1.55$), combined with Poisson noise in the phase-space distribution of stars at the time of gas expulsion and with the stochasticity characterizing the expansion phase due to the relatively small number of massive stars staying bound to the cluster surviving core.

 To be consistent with the other works where the violent relaxation is neglected we introduce the cluster ``initial'' mass as its mass at the end of violent relaxation, when the cluster stops to lose mass violently due to gas removal.
 In the scope of this study we have provided two cluster ``initial'' mass estimates, the Jacobi mass and the ``extended'' mass.
 The Jacobi mass is the stellar mass within one Jacobi radius, $R_J$, while the extended mass is the stellar mass within two Jacobi radii, $2R_J$.
 We have estimated the latter, because escaping stars form an envelope around their natal cluster, which stays in the cluster surroundings for a few tens of Myrs (see Fig.~\ref{fig:densevol}).
 Such an envelope possibly contribute to the mass estimate of extra-galactic young star clusters, where the membership analysis is impossible and the measurement of the tidal radius is not straightforward.

 From our simulations we have found that star clusters formed with a high global SFE ($\geq0.20$) dissolve in a tight mass-dependent regime ($t_{dis}\propto M_{init}^{0.7}$), in agreement with earlier works (Fig.~\ref{fig:lifemass}).
 In the solar neighborhood, these clusters dissolve in a way similar to the empirical relation of \citet{BL03}, with a dissolution time for a $10^4\msol$ cluster of $t_4^{dis}= 3.8\text{ Gyr}$.
 This is almost a factor of two shorter than the estimate of \citet{BM03} ($t_4^{dis}= 6.9\text{ Gyr}$), but, still longer than that given in \citet{Lamers+05a} ($t_4^{dis}= 1.3\pm0.5\text{ Gyr}$).
 
 In contrast, model clusters formed with $SFE_{gl}=0.15$ dissolve more quickly than high SFE clusters ($t_4^{dis}\sim 1.3\text{ Gyr}$), present a shallower MDD relation ($t_{dis}\propto M_{init}^{0.5}$), and their dissolution time is affected by a relatively large scatter.
 That is, variations of the bound mass fraction at the end of violent relaxation (due to the stochastic impact of the massive stars described above) can modify sensitively the cluster dissolution time. 
 The lower the bound mass fraction, the shorter the dissolution time (Fig.~\ref{fig:tdisFb}).
 We have found that some of our $SFE_{gl}=0.15$ model clusters can dissolve within 100-200 Myr, while the high-SFE clusters with the same ``initial'' mass can live longer than a Gyr (e.g. consider the vertical bin embracing a cluster mass range from $2\times10^3\msol$ to $4\times10^3\msol$ in the right panels of Fig.~\ref{fig:lifemass}).
 
 Nevertheless, taken all together, our model clusters follow an MDD relation, albeit with a relatively large scatter.
 The relation between the dissolution time and cluster initial mass becomes close to that observationally found by \citet{Lamers+05a} for the solar neighborhood if the cluster population is dominated by low-SFE ($SFE_{gl}=0.15$) clusters.
 Such an assumption is a real possibility for the solar neighborhood since nearby embedded clusters usually show low SFEs.
 Therefore, in this study, we propose to  {recover the cluster dissolution time for the solar neighborhood found by \citet{Lamers+05a} in a way that is alternative} to that proposed by \citet{Gieles+06}.
 While \citet{Gieles+06} propose that cluster dissolution time inferred from observations by \citet{Lamers+05a} is shorter than the prediction of \citet{BM03} due to the additional destructive processes associated to GMC encounters, we propose here that the difference between theory and observations can be removed once one takes into account diffuse weakly bound star clusters, arising from low $SFE_{gl}=0.15$.

 The distribution of the dissolution times of high-SFE ($SFE_{gl}\geq0.17$) and low-SFE ($SFE_{gl}=0.15$) clusters is provided in Fig.~\ref{fig:tdishist}. It shows that our low-SFE clusters usually dissolve within 1~Gyr and our high-SFE clusters, in contrast, usually survive beyond a Gyr.

 If we consider our low-SFE clusters only, estimate their initial mass as the extended mass at $t=70$~Myr and neglect the low-mass end (as for extra-galactic observations, the unbound envelope-stars usually contribute to the cluster mass and the low mass clusters often remain undetectable), an interesting apparent MID relation can emerge (see the flat part of red curve with $ 10^3\msol < M_{2J} < 10^4\msol$ in Fig.~\ref{fig:meantdis}). 
 But since our simulations do not cover that high a range of cluster ``initial'' masses, we cannot firmly argue about MID based on our current data-set.
 

We have found a correlation between the cluster life expectancy and the volume density contrast between cluster center and  outskirts (top panels of Fig. \ref{fig:rho}). 
 The higher the density contrast, the longer the cluster lives.
 We have found that they also correlate with cluster initial mass, which is shown with nice patterns in the bottom panels of Fig.~\ref{fig:rho}.
 
 All correlations found in this paper are tighter if we consider them at the age of  $t=70$~Myr rather than at $t=30$~Myr.
 Therefore,  we propose that for cluster lifetime studies, it is more appropriate to measure cluster ``initial'' parameters as mass, central density and structure about 70~Myr after gas expulsion, when star clusters have mostly re-virialized and are cleaned up from their expanding stellar envelope. 

 So we summarize that low-mass clusters are unable to live for as long as their high-mass counterparts.
 However, high-mass clusters can easily dissolve on a short time, if formed with a small SFE, but still outlive most of their low-mass counterparts. 
 Overall, our model clusters dissolve in a mass-dependent regime, although with different dissolution times for low- and high-SFE models.
 For low-SFE ($SFE_{gl}=0.15$), an apparent MID mode can emerge if we define the extended mass as the initial mass of clusters and if we restrict our attention to clusters more massive than $10^3\msol$, as is often the case in extra-galactic studies.
In a forthcoming paper, we will consider how our simulations respond to environmental variations.

\acknowledgments
 {We thank the referee for his/her helpful comments.}

This work was supported by Sonderforschungsbereich SFB 881 ``The Milky Way System'' (subproject B2) of the German Research Foundation (DFG). 
The authors gratefully acknowledge Prof. Dr. Walter Dehnen for his support and discussions in connection with implementing the code \textsc{mkhalo} for our purposes.
BS gratefully acknowledges Prof. Dr. Rainer Spurzem for his support with accessing the high-performance computing clusters as JURECA and LAOHU.
BS acknowledges the support within program No. BR05236322 funded by the Ministry of Education and Science of the Republic of Kazakhstan.
PB acknowledges support by the Chinese Academy of Sciences through the
Silk Road Project at NAOC, through the ``Qianren'' special foreign experts
program, and the President's International Fellowship for Visiting
Scientists program of CAS and also the Strategic Priority Research Program
(Pilot B) ``Multi-wavelength gravitational wave universe'' of the Chinese Academy
of Sciences (No. XDB23040100).
PB acknowledges the support of the Volkswagen Foundation under the Trilateral
Partnerships grant No. 90411 and the special support by the NASU under
the Main Astronomical Observatory GRID/GPU computing cluster project.
This work benefited from support by the International Space Science Institute, Bern, Switzerland,  through its International Team program ref. no. 393 ``The Evolution of Rich Stellar Populations \& BH Binaries'' (2017-18).
The authors gratefully acknowledge the computing time granted by the John von Neumann Institute for Computing (NIC) and provided on supercomputer JURECA at J\"ulich Supercomputing Center (JSC) through grant HHD28.
The authors acknowledge support by the state of Baden-W\"urttemberg through bwHPC (bwForCluster MLS\&WISO) and the German Research Foundation (DFG) through grant INST 35/1134-1 FUGG.
We also use the smaller GPU cluster KEPLER, funded under the grants I/80 041-043 and
I/81 396 of the Volkswagen Foundation, and the special GPU accelerated supercomputer LAOHU at the Center of Information and Computing at National Astronomical Observatories, Chinese Academy of Sciences, funded by the Ministry of Finance of the People's Republic of China under the grant ZDYZ2008-2.
\bibliographystyle{aasjournal}
\bibliography{paper2}

\end{document}